\documentclass[prc,superscriptaddress,unsortedaddress,twocolumn,showpacs,preprintnumbers,amsmath,amssymb,floatfix,dvipdfmx]{revtex4}
\usepackage[dvipdfmx]{graphicx}
\usepackage{amsmath}
\usepackage{amssymb}
\usepackage{mathrsfs}
\usepackage{multirow}
\usepackage{bm}
\usepackage{color}

\def\la{\mathrel{\mathpalette\fun <}}
\def\ga{\mathrel{\mathpalette\fun >}}
\def\fun#1#2{\lower3.6pt\vbox{\baselineskip0pt\lineskip.9pt
\ialign{$\mathsurround=0pt#1\hfil##\hfil$\crcr#2\crcr\sim\crcr}}}

\newcommand{\beq}{\begin{equation}}
\newcommand{\eeq}{\end{equation}}
\newcommand{\bea}{\begin{eqnarray}}
\newcommand{\eea}{\end{eqnarray}}

\newcommand{\bfi}[1]{\mbox{\boldmath $#1$}}

\newcommand{\vK}{{\bfi K}}

\newcommand{\vs}{{\bfi s}}

\newcommand{\vrr}{{\bfi r}}
\newcommand{\vR}{{\bfi R}}

\begin{document}


\title{Microscopic optical potentials for $^{4}$He scattering}

\author{Kei Egashira}
\email[]{egashira@phys.kyushu-u.ac.jp}
\affiliation{Department of Physics, Kyushu University, Fukuoka 812-8581, Japan}

\author{Kosho Minomo}
\email[]{minomo@rcnp.osaka-u.ac.jp}
\affiliation{Research Center for Nuclear Physics, Osaka University, Ibaraki 567-0047, Japan}

\author{Masakazu Toyokawa}
\email[]{toyokawa@phys.kyushu-u.ac.jp}
\affiliation{Department of Physics, Kyushu University, Fukuoka 812-8581, Japan}

\author{Takuma Matsumoto}
\email[]{matsumoto@phys.kyushu-u.ac.jp}
\affiliation{Department of Physics, Kyushu University, Fukuoka 812-8581, Japan}

\author{Masanobu Yahiro}
\email[]{yahiro@phys.kyushu-u.ac.jp}
\affiliation{Department of Physics, Kyushu University, Fukuoka 812-8581, Japan}

\date{\today}

\begin{abstract}
We present a reliable double-folding (DF) model for $^{4}$He-nucleus 
scattering, using the Melbourne $g$-matrix nucleon-nucleon interaction 
that explains nucleon-nucleus scattering with no adjustable parameter. 
In the DF model, only the target density is taken as the local density 
in the Melbourne $g$-matrix. 
For $^{4}$He elastic scattering from $^{58}$Ni and $^{208}$Pb targets 
in a wide range of incident energies from 20~MeV/nucleon to 200~MeV/nucleon, 
the DF model with the target-density approximation (TDA) yields much 
better agreement with the experimental data than the usual DF model 
with the frozen-density approximation in 
which the sum of projectile and target densities is taken 
as the local density. 
We also discuss the relation between the DF model with the TDA 
and the conventional folding model 
in which the nucleon-nucleus potential is folded 
with the $^{4}$He density. 
\end{abstract}


\maketitle
\section{Introduction}
\label{Introduction}

Microscopic derivation of nucleon-nucleus (NA) and 
nucleus-nucleus (AA) optical potentials is a goal of nuclear reaction theory. 
The optical potential is an important quantity to describe not only the
elastic scattering but also more complicated reactions such as inelastic
scattering, breakup and transfer reactions. For the latter case, the
optical potential is used as a key input in theoretical calculations
such as the distorted-wave Born approximation and the continuum
discretized coupled-channels
method~\cite{CDCC-review1,CDCC-review2,Yahiro:2012tk}.

The $g$-matrix folding model is a powerful tool of deriving  
NA and AA optical potentials. In the model, the optical potential is 
calculated by folding the $g$-matrix effective nucleon-nucleon (NN)
interaction~\cite{M3Y,JLM,Brieva-Rook,Satchler-1979,Satchler,CEG,Rikus-von
Geramb,Amos,CEG07,Saliem}  with the target density for NA scattering 
and the projectile and target densities for AA scattering; 
see for example
Refs.~\cite{DFM-standard-form,Arellano:1995,rainbow,DFM-standard-form-2,Sum12}  
for the folding procedure. 
The folding model for NA and AA scattering are referred to as the
single-folding model and the double-folding (DF) model, respectively.
For NA elastic scattering, the model is quite
successful in reproducing the 
experimental data systematically with no free parameter, 
when the Melbourne $g$-matrix~\cite{Amos} is 
used as an effective NN interaction in the folding calculations.  
As an important advantage of the $g$-matrix folding model, the model
takes account of nuclear medium effects.    
The $g$-matrix is calculated in nuclear matter and hence 
depends on nuclear-matter density $\rho$. When the optical potential is 
evaluated from the $g$-matrix in the folding procedure, 
the nuclear-matter density is replaced by the target density at the location 
of interacting nucleon pair.
This approximation is called the local-density approximation.

The NA potential thus derived is non-local and 
thereby not so practical in many applications. 
It is, however, possible to localize the potential with the Brieva-Rook 
approximation \cite{Brieva-Rook}. Recently the validity of the
approximation was shown in Ref. \cite{Minomo:2009ds,Hag06}. In fact, the
local version of $g$-matrix folding potential describes NA scattering
with no adjustable parameter~\cite{Toyokawa:2013uua}, 
and close to the phenomenological NA optical
potentials~\cite{Perey-Perey,Dirac1,Dirac2,Koning-Delaroche}.

From a theoretical viewpoint based on the multiple scattering 
theory~\cite{Watson,KMT,Yahiro-Glauber}, 
the multiple NN scattering series 
in AA collision~\cite{Yahiro-Glauber} is more complicated than that in 
NA collision~\cite{Watson,KMT}. 
In this sense, microscopic understanding of the optical potentials  
is relatively more difficult for AA scattering than for NA scattering. 
One of the simplest composite projectiles is $^{4}$He, since it is
almost inert. For $^{4}$He-nucleus elastic scattering, a systematic
analysis was made \cite{Furumoto:2006ek} by using the $g$-matrix
interaction proposed by Jeukenne, Lejeune, and Mahaux (JLM) \cite{JLM}.  
The JLM $g$-matrix folding model reproduces 
measured differential cross sections 
for $^{4}$He elastic scattering at incident energies ranging 
from 10 to 60 MeV/nucleon, if 
the real and imaginary parts of the folding potential are reduced by about 25\% and 35\%, respectively.
In the JLM $g$-matrix, nuclear medium effects 
are included  only partly, so that the normalization factors are always 
introduced. This fact strongly suggests that the parameter-free analysis 
based on the Melbourne $g$-matrix folding model should be made 
for $^{4}$He-nucleus elastic scattering.

In the DF procedure, 
the frozen-density approximation (FDA) is usually taken. Namely, 
one takes as the local density 
the sum of projectile and target densities, $\rho_{\rm P}$ and $\rho_{\rm T}$, 
at the midpoint of interacting two nucleons, one in projectile (P) and the 
other in target (T):
\bea
g(\rho)=g({\rho_{\rm P}}+{\rho_{\rm T}}) . 
\label{FD-approx}
\eea
Very recently, the Melbourne $g$-matrix folding 
model with the FDA was applied to $^{12}$C+$^{12}$C and $^{20-32}$Ne+$^{12}$C 
elastic scattering at intermediate energies with success in
reproducing measured  
differential cross sections $d \sigma/d \Omega$ 
and total reaction cross sections $\sigma_{\rm R}$ 
with no free parameter~\cite{Min11,Min12,Sum12}.
In the calculations, the densities of unstable nuclei $^{20-32}$Ne 
were evaluated by antisymmetrized molecular 
dynamics (AMD)~\cite{Kimura,Kimura1} 
with the Gogny-D1S interaction~\cite{GognyD1S}. 
The AMD wave functions successfully describe low-lying
spectra of Ne isotopes~\cite{Kimura}. 
The microscopic approach concluded that 
$^{30-32}$Ne in the ``island of inversion'' have large deformation and 
$^{31}$Ne has a deformed halo structure~\cite{Min11,Min12,Sum12}. 
This indicates that the $N=20$ magicity disappears. 
The Melbourne $g$-matrix folding model is thus a powerful tool of 
not only understanding the reaction mechanism but also determining 
the structure of unstable nuclei.

In this paper, we microscopically describe $^{4}$He elastic scattering 
from heavier targets such as $^{58}$Ni and $^{208}$Pb 
in a wide range of incident energies from 20~MeV/nucleon to 200~MeV/nucleon, 
using the Melbourne $g$-matrix DF model with no adjustable parameter.
After showing that the DF model with the FDA cannot reproduce 
measured $d \sigma/d \Omega$ and $\sigma_{\rm R}$ for the scattering, 
we propose a new approximation instead of the FDA. 
In the approximation, 
only the target density is taken as the local density. 
This approximation is referred to as the target-density approximation 
(TDA) in this paper. 
The reliability of the TDA is shown theoretically with 
the multiple scattering theory \cite{Watson,KMT,Yahiro-Glauber} 
and phenomenologically by showing that 
the DF model with the TDA well reproduces 
the data on $d \sigma/d \Omega$ and $\sigma_{\rm R}$. 
We also investigate the reliability of the conventional 
nucleon-nucleus folding (NAF) model in which the NA potential 
is folded with the $^{4}$He density.

In Sec. \ref{Theoretical framework}, we recapitulate 
the Melbourne $g$-matrix DF model and 
show the reliability of the TDA theoretically. 
Numerical results are shown in Sec.~\ref{Results}. 
Section \ref{Summary} is devoted to a summary.

\section{Model building}
\label{Theoretical framework}

AA scattering can be described by
the many-body Schr\"odinger equation,
\bea
(T_R +h_{\rm P}+h_{\rm T}+ \sum_{i \in {\rm P}, j
\in {\rm T}} v_{ij}-E){\Psi}^{(+)}=0,
\label{schrodinger-bare}
\eea
with the realistic NN interaction $v_{ij}$,
where $T_R$ stands for the kinetic energy with respect to the relative
coordinate ($\vR$) between the projectile (P) and the target (T), 
$E$ is the total energy and 
$h_{\rm P}$ ($h_{\rm T}$) means the internal Hamiltonian of P (T).
Using the multiple scattering theory~\cite{Watson, KMT} for 
AA scattering~\cite{Yahiro-Glauber}, one can rewrite 
Eq.~\eqref{schrodinger-bare} into
\bea
(T_R +h_{\rm P}+h_{\rm T}+ \sum_{i \in {\rm P}, j \in {\rm T}}
\tau_{ij}-E){\hat \Psi}^{(+)}=0 
\label{schrodinger-effective}
\eea
with the effective NN interaction $\tau_{ij}$ 
defined by 
\bea
\tau_{ij}=v_{ij}+v_{ij}G_0\tau_{ij}
\label{tau}
\eea
with 
\bea
G_0=\frac{{\cal P_{\rm P}}{\cal P_{\rm T}}}{E-K-h_{\rm P}-h_{\rm T}+i\epsilon},
\label{D0-AA}
\eea
where
${\cal P_{\rm P}}$ (${\cal P_{\rm T}}$) denotes the projection operator onto 
the space of antisymmetrized wave functions of P (T).
In the derivation of Eq. \eqref{schrodinger-effective}, the 
antisymmetrization between nucleons in P and those in T has been neglected, 
but it is shown in Refs.~\cite{Takeda,Picklesimer} that 
the antisymmetrization effects are well taken care of 
by using $\tau_{ij}$ that is properly symmetrical with respect to
the exchange of the colliding nucleons. 
Since the effective NN interaction $\tau_{ij}$ includes 
nuclear medium effects, the $g$-matrix ($g_{ij}$) is often used as 
such $\tau_{ij}$~\cite{M3Y,JLM,Brieva-Rook,Satchler-1979,Satchler,CEG,
Rikus-von Geramb,Amos,CEG07,rainbow}.

Since $g_{ij}$ also includes projectile- and target-excitation effects 
approximately, Eq.~\eqref{schrodinger-effective} can be further rewritten 
into the single-channel equation
\bea
[T_R+U-E_{\rm in}]\psi=0,
\eea
with the folding potential
\bea
U(\vR)=\langle \Phi_0 | \sum_{i \in {\rm P}, j \in {\rm T}}
g_{ij} | \Phi_0 \rangle \;,
\eea
where the incident energy $E_{\rm in}$ is related to the total energy $E$ 
as $E=E_{\rm in}+e_0({\rm P})+e_0({\rm T})$ for the grand-state energies, 
$e_0(\rm P)$ and $e_0(\rm T)$, of P and T. The wave function $\Phi_0$ 
denotes the product of the ground states of P and T, while 
$\psi$ means the relative wave function between P and T.
This is nothing but the $g$-matrix DF model. 
In the actual calculations, 
the FDA shown in Eq.~\eqref{FD-approx} is usually taken and 
the Coulomb potential $U_{\rm Coul}$ is added to the resulting $U$.

Now we consider $^{4}$He scattering from heavier nuclei. 
In the scattering, the projectile ($^{4}$He) is hardly excited, 
whereas the target is excited easily. 
As a good approximation we can hence neglect projectile excitations. Namely, 
we can replace $h_{\rm P}$ by the ground-state energy $e_0({\rm P})$ and hence 
${\cal P_{\rm P}}{\cal P_{\rm T}}$ by ${\cal P_{\rm T}}$: 
\bea
G_0 \approx \frac{{\cal P_{\rm T}}}{E-K-e_0({\rm P})-h_{\rm T}+i\epsilon} .
\label{D0-AA-2}
\eea
After the approximation, the $\tau_{ij}$ includes nuclear medium 
effects from T, but not from P. We should therefore replace $\tau_{ij}$ by 
the $g$-matrix depending only on $\rho_{\rm T}$: 
\bea
g(\rho)=g({\rho_{\rm T}}) . 
\label{TD-approx}
\eea
This is the TDA proposed in the present paper. 
The reliability of the TDA is confirmed also phenomenologically 
in Sec. \ref{Results} by comparing the theoretical results 
with the experimental data and showing that 
the TDA is much better than the FDA.

Next we recapitulate the single folding model
for NA scattering and the DF model for AA scattering. 
As for the detail of the models, for example, 
see Refs. \cite{Brieva-Rook,CEG07,Saliem,Arellano:1995,Minomo:2009ds,Toyokawa:2013uua} 
for NA scattering 
and Refs.~\cite{CEG07,DFM-standard-form,DFM-standard-form-2,Hag06,Sum12} 
for AA scattering. 
The DF potential $U=V+iW$ 
consists of the direct and exchange parts,
$U^{\rm DR}$ and $U^{\rm EX}$
~\cite{DFM-standard-form,DFM-standard-form-2}: 
\bea
U(\vR)=U^{\rm DR}(\vR)+U^{\rm EX}(\vR)+U_{\rm Coul}(\vR) 
\eea
with 
\bea
\label{eq:UD}
U^{\rm DR}(\vR) \hspace*{-0.15cm} &=& \hspace*{-0.15cm}
\sum_{\mu,\nu}\int \rho^{\mu}_{\rm P}(\vrr_{\rm P})
            \rho^{\nu}_{\rm T}(\vrr_{\rm T})
            g^{\rm DR}_{\mu\nu}(s;\rho_{\mu\nu}) d \vrr_{\rm P} d \vrr_{\rm T}, \\
\label{eq:UEX}
U^{\rm EX}(\vR) \hspace*{-0.15cm} &=& \hspace*{-0.15cm}\sum_{\mu,\nu}
\int \rho^{\mu}_{\rm P}(\vrr_{\rm P},\vrr_{\rm P}-\vs)
\rho^{\nu}_{\rm T}(\vrr_{\rm T},\vrr_{\rm T}+\vs) \nonumber \\
            &&~~\hspace*{-1.0cm}\times g^{\rm EX}_{\mu\nu}(s;\rho_{\mu\nu}) \exp{[-i\vK(\vR) \cdot \vs/M]}
            d \vrr_{\rm P} d \vrr_{\rm T},~~~~
\eea
where $\vrr_{\rm P}$ ($\vrr_{\rm T}$) stands for the coordinate of 
the interacting nucleon from the center of mass of P (T), 
$\vs=\vrr_{\rm P}-\vrr_{\rm T}+\vR$, 
and each of $\mu$ and $\nu$ denotes the $z$-component of isospin. 
Here $\rho^{\mu}_{\rm P}(\vrr_{\rm P})$ and $\rho^{\nu}_{\rm T}(\vrr_{\rm T})$ 
are one-body densities of P and T and 
$\rho^{\mu}_{\rm P}(\vrr_{\rm P},\vrr_{\rm P}-\vs)$ and 
$\rho^{\nu}_{\rm T}(\vrr_{\rm T},\vrr_{\rm T}+\vs)$ are 
mixed densities of P and T, respectively. 
The non-local $U^{\rm EX}$ has been localized in Eq.~\eqref{eq:UEX}
with the local semi-classical approximation~\cite{Brieva-Rook}, 
where the local momentum $\hbar \vK(\vR)$ of P relative to T is 
defined by $\hbar K(R) \equiv \sqrt{2\mu_{\rm PT} (E_{\rm in} -U(R))}$ 
with the reduced mass $\mu_{\rm PT}$ between P and T, 
and $M=A_{\rm P} A_{\rm T}/(A_{\rm P} +A_{\rm T})$ 
for the mass numbers, $A_{\rm P}$ and $A_{\rm T}$, of P and T. 
The validity of the localization is shown in 
Refs.~\cite{Hag06,Minomo:2009ds}. 
The direct and exchange parts, $g^{\rm DR}_{\mu\nu}$ and
$g^{\rm EX}_{\mu\nu}$, of the $g$-matrix depend on the local density
at the midpoint of the interacting nucleon pair:
\bea
 \rho_{\mu\nu}=\rho^{\mu}_{\rm P}(\vrr_{\rm P}-\vs/2)
 +\rho^{\nu}_{\rm T}(\vrr_{\rm T}+\vs/2)
\label{local-density approximation-1}
\eea
in the FDA and 
\bea
 \rho_{\mu\nu}=\rho^{\nu}_{\rm T}(\vrr_{\rm T}+\vs/2)
\label{local-density approximation-2}
\eea
in the TDA; see Ref.~\cite{Sum12} 
for the explicit forms of $g^{\rm DR}_{\mu\nu}$ and 
$g^{\rm EX}_{\mu\nu}$.

We now consider NA scattering at an incident energy 
$E_{\rm in}^{\rm N}$. 
The single folding potentials $U_{\mu}=V_{\mu}+iW_{\mu}$ 
for proton ($\mu=-1/2$) and neutron ($\mu=1/2$) scattering are also 
composed of $U_{\mu}^{\rm DR}$ and $U_{\mu}^{\rm EX}$:
\bea
U_{\mu}(\vrr_{\mu})=U_{\mu}^{\rm DR}(\vrr_{\mu})+U_{\mu}^{\rm EX}(\vrr_{\mu})
+U_{\rm Coul}(\vrr_{\mu}) 
\eea
with 
\bea
\label{eq:UD-NA}
U_{\mu}^{\rm DR}(\vrr_{\mu}) \hspace*{-0.15cm} &=& \hspace*{-0.15cm}
\sum_{\nu}\int 
            \rho^{\nu}_{\rm T}(\vrr_{\rm T})
            g^{\rm DR}_{\mu\nu}(s;\rho_{\mu\nu}) d \vrr_{\rm T}, 
\\ \label{eq:UEX-NA}
U_{\mu}^{\rm EX}(\vrr_{\mu}) \hspace*{-0.15cm} &=& \hspace*{-0.15cm}\sum_{\nu}
\int 
\rho^{\nu}_{\rm T}(\vrr_{\rm T},\vrr_{\rm T}+\vs) \nonumber \\
            &&~~\hspace*{-1.0cm}\times g^{\rm EX}_{\mu\nu}(s;\rho_{\mu\nu}) 
\exp{[-i\vK_{\mu}(\vrr_{\mu}) \cdot \vs]}
            d \vrr_{\rm T},~~~~
\eea
where $\vs=\vrr_{\mu}-\vrr_{\rm T}$ for $\vrr_{\mu}$ the coordinate 
of an incident nucleon from the center of mass of T, the local 
density $\rho_{\mu\nu}$ is obtained by 
Eq. \eqref{local-density approximation-2}, and 
the local momentum $\hbar \vK_{\mu}(\vrr_{\mu})$ 
between the incident nucleon (N) and T is defined by 
$\hbar K_{\mu}(r_{\mu}) \equiv \sqrt{2\mu_{\rm NT} 
(E_{\rm in}^{\rm N} -U_{\mu}(r_{\mu}))}$ for 
the reduced mass $\mu_{\rm NT}$ between N and T.

When AA scattering at high $E_{\rm in}$ is compared with NA scattering 
at $E_{\rm in}^{\rm N}=E_{\rm in}/A_{\rm P}$ for heavy targets satisfying 
$A_{\rm T}\gg A_{\rm P}>1$, the local momenta $\hbar \vK_{\mu}(\vrr_{\mu})$ 
and $\hbar \vK(\vR)$ nearly agree with their asymptotic values 
$\hbar \vK_{\mu}(\infty)$ and $\hbar \vK(\infty)$, respectively, so that 
\bea
\vK_{\mu}(\infty)=\vK(\infty)/M .
\label{K-relation}
\eea
Taking the relation \eqref{K-relation} and the TDA, 
one can get
\bea
\label{eq:UD-2}
U^{\rm DR}(\vR) \hspace*{-0.15cm} &\approx& \hspace*{-0.15cm}
\sum_{\mu} \int \rho^{\mu}_{\rm P}(\vrr_{\rm P}) 
U_{\mu}^{\rm DR}(\vR+\vrr_{\rm P})
             d \vrr_{\rm P}, \\
\label{eq:UEX-2}
U^{\rm EX}(\vR) \hspace*{-0.15cm} &\approx& \hspace*{-0.15cm}\sum_{\mu}
\int \rho^{\mu}_{\rm P}(\vrr_{\rm P}) 
U_{\mu}^{\rm EX}(\vR+\vrr_{\rm P}) d \vrr_{\rm P}.~~~~
\eea
In the derivation of Eq.~\eqref{eq:UEX-2}, we have also used the approximation 
$\rho^{\mu}_{\rm P}(\vrr_{\rm P},\vrr_{\rm P}-\vs) 
\approx \rho^{\mu}_{\rm P}(\vrr_{\rm P},\vrr_{\rm P}) = 
\rho^{\mu}_{\rm P}(\vrr_{\rm P})$ good 
in the peripheral region of T that is important for the elastic 
scattering~\cite{Minomo:2009ds}. 
For $^{4}$He scattering from heavier targets at high $E_{\rm in}$, 
the DF potential $U^{\rm DR}+U^{\rm EX}$ with 
the TDA is thus obtained with reasonable accuracy 
by folding the nucleon-nucleus potential $U_{\mu}^{\rm DR}+U_{\mu}^{\rm EX}$ 
with the projectile density $\rho^{\mu}_{\rm P}$. 
This is the NAF model mentioned in Sec. \ref{Introduction}. 
This model is quite practical, 
since one can use the phenomenological NA optical potential 
instead of $U_{\mu}$. 
The validity of this model is also investigated later 
in Sec. \ref{Results}. The condition that the local momenta 
$\hbar \vK_{\mu}(\vrr_{\mu})$ and $\hbar \vK(\vR)$ are close to 
their asymptotic values is well satisfied at large $R$, even if $E_{\rm in}$ 
is small. Since $^{4}$He scattering from heavy targets are quite peripheral 
at small $E_{\rm in}$, one can expect that 
the NAF model is good also for small $E_{\rm in}$. 
This is also discussed in Sec. \ref{Results}.

\section{Results}
\label{Results}

We analyze measured $d \sigma/d \Omega$ and $\sigma_{\rm R}$ for 
$^{4}$He elastic scattering from $^{58}$Ni and $^{208}$Pb 
targets in the region $20 \la E_{\rm in}/A_{\rm P} \la 200$~MeV, 
using the following three models: 
\begin{enumerate}
 \item The DF model with the TDA (the DF-TDA model)
 \item The DF model with the FDA (the DF-FDA model)
 \item The NAF model
\end{enumerate}
As the $^{4}$He density $\rho_{\rm P}$, 
we use the phenomenological proton-density~\cite{C12-density} 
determined from the electron scattering 
in which the finite-size effect due to the proton charge 
is unfolded in the standard manner~\cite{Singhal}. 
The neutron density is assumed to have the same geometry as the proton one. 
As the target density $\rho_{\rm T}$, we take the matter densities calculated 
by the spherical Hartree-Fock (HF) model 
with the Gogny-D1S interaction~\cite{GognyD1S} in which 
the spurious center-of-mass motion is removed 
in the standard manner~\cite{Sum12}.

\begin{figure}[htbp]
\begin{center}
 \includegraphics[width=0.40\textwidth,clip]{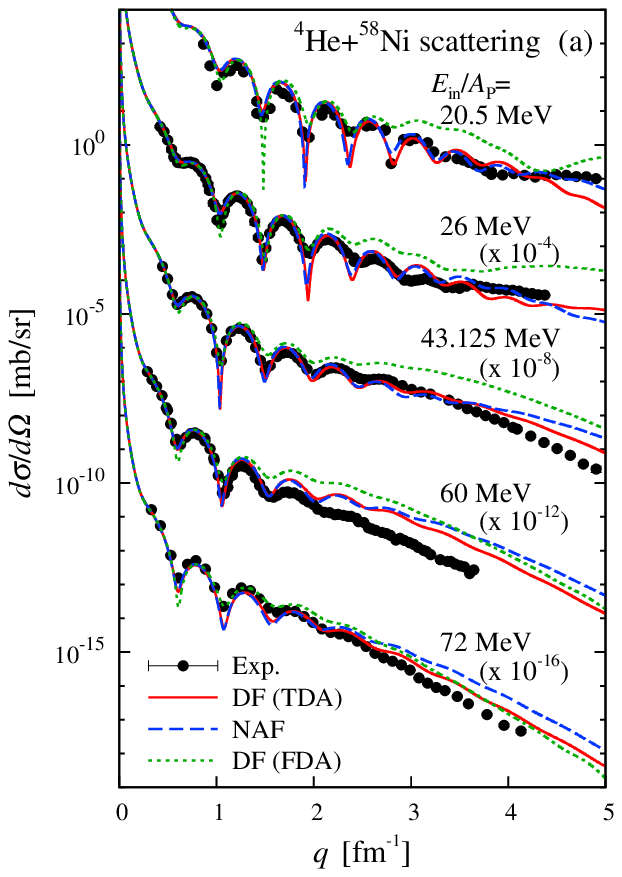}
 \includegraphics[width=0.40\textwidth,clip]{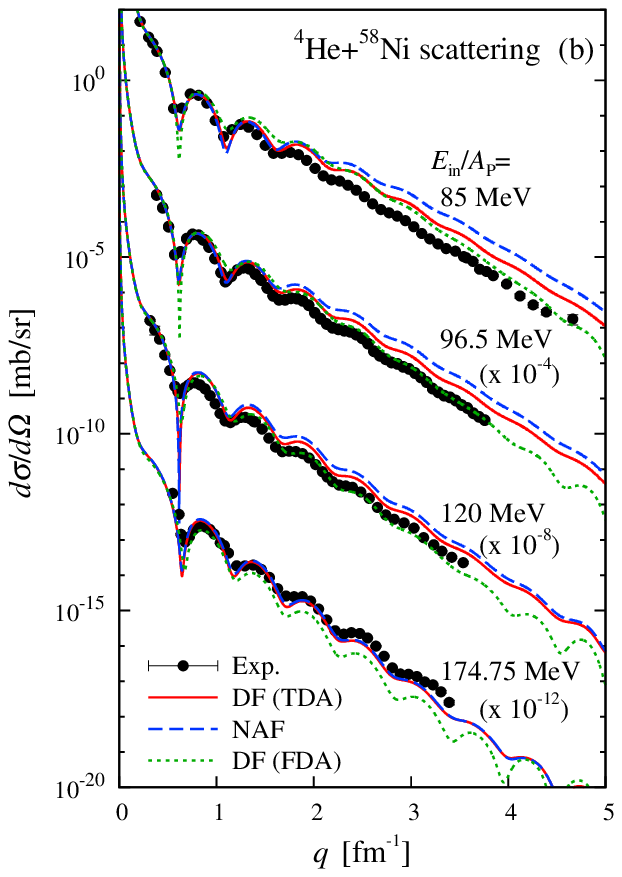}
 \caption{(Color online) 
Differential cross sections as a function of transfer momentum $q$ 
for $^{4}$He+$^{58}$Ni elastic scattering at 
(a) $E_{\rm in}/A_{\rm P}=20\mbox{--}72$~MeV and 
(b) $E_{\rm in}/A_{\rm P}=85\mbox{--}175$~MeV.
The cross section at each $E_{\rm in}/A_{\rm P}$ is multiplied by 
the factor shown in the panel. 
The solid (dotted) line stands 
the results of the DF-TDA (DF-FDA) model, 
whereas the dashed line denotes the results of the NAF model. 
The experimental data are taken from Refs.~~\cite{E21:Chang:1976,E26:Rebel:1972,E43:Albinski:1985,E60:Lui:2006,E72-175:Bonin:1985,E97:Nayak:2006}.
}
 \label{Fig-XSEC-He4Ni58}
\end{center}
\end{figure}

\begin{figure}[htbp]
\begin{center}
 \includegraphics[width=0.35\textwidth,clip]{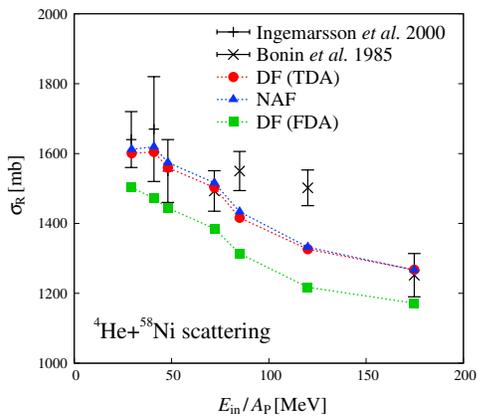}
 \caption{(Color online) Total reaction cross section  $\sigma_{\rm R}$
as a function of $E_{\rm in}/A_{\rm P}$ for $^{4}$He+$^{58}$Ni scattering 
at  $E_{\rm in}/A_{\rm P}= 20\mbox{--}175$~MeV. 
The circles (squares) stand 
the results of the DF-TDA (DF-FDA) model, 
whereas the triangles denote the results of the NAF model. 
The experimental data are taken from \cite{E29-48:Ingemarsson:2000,E72-175:Bonin:1985}.
}
 \label{Fig-ReactionXSEC-He4Ni58}
\end{center}
\end{figure}

Figure \ref{Fig-XSEC-He4Ni58} shows $d\sigma/d\Omega$ 
as a function of transfer momentum $q$ for 
$^{4}$He+$^{58}$Ni scattering in 
$E_{\rm in}/A_{\rm P}=20\mbox{--}175$~MeV. 
For lower incident energies of $E_{\rm in}/A_{\rm P}=20\mbox{--}60$~MeV, 
the DF-FDA model (dotted line) overestimates 
the experimental data \cite{E21:Chang:1976,E26:Rebel:1972,E43:Albinski:1985,E60:Lui:2006}, but this problem is solved by 
the DF-TDA model (solid line) that well reproduces the data. 
For higher energies around $E_{\rm in}/A_{\rm P}=175$~MeV, 
meanwhile, the DF-FDA model underestimates 
the experimental data \cite{E72-175:Bonin:1985}, but this problem is also 
solved by the DF-TDA model that reproduces the data. 
For intermediate energies of $E_{\rm in}/A_{\rm P}=72\mbox{--}120$~MeV, 
the difference between the DF-TDA and DF-FDA results is rather small, 
so that both the models reasonably reproduce the data. 
In great detail, for $E_{\rm in}/A_{\rm P}=85$~MeV, the DF-TDA result 
is better than the DF-FDA result at $q \la 2$~fm$^{-1}$, whereas 
the latter is better than the former at $q \ga 3$~fm$^{-1}$. 
For $E_{\rm in}/A_{\rm P}=97$~MeV, the DF-FDA model is slightly better 
than the DF-TDA model, but this seems to be accidental, 
since for $\sigma_{\rm R}$  
the DF-TDA model (circles) yields better agreement with the data \cite{E29-48:Ingemarsson:2000,E72-175:Bonin:1985} 
than the DF-FDA model (squares) 
as shown in Fig. \ref{Fig-ReactionXSEC-He4Ni58}. 
Throughout these analyses, we can conclude that 
the DF-TDA model is much better than the DF-FDA model.

Next we compare the DF-TDA model with the NAF model in Figs. 
\ref{Fig-XSEC-He4Ni58} and \ref{Fig-ReactionXSEC-He4Ni58}. 
For $\sigma_{\rm R}$, 
the NAF model (triangles) well 
simulates the DF-TDA result (circles) and hence yields 
much better agreement with the data \cite{E29-48:Ingemarsson:2000,E72-175:Bonin:1985} 
than the DF-FDA model (squares). 
For $d \sigma/d \Omega$ at higher energies of 
$E_{\rm in}/A_{\rm P}=120\mbox{--}175$~MeV, as expected, 
the NAF results (dashed lines) well reproduce 
the DF-TDA results (solid lines). 
Also for lower energies of $E_{\rm in}/A_{\rm P}=20\mbox{--}43$~MeV, 
the NAF model well simulates the DF-TDA results, since 
the elastic scattering is quite peripheral, as shown below. 
For intermediate energies of $E_{\rm in}/A_{\rm P}=60\mbox{--}97$~MeV, 
however, the NAF results are deviated sizably from the DF-TDA results.

\begin{figure}[htbp]
\begin{center}
 \includegraphics[width=0.35\textwidth,clip]{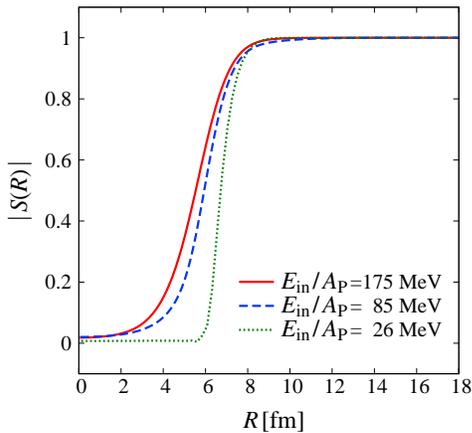}
 \caption{(Color online) 
$R$ dependence of the absolute value of the elastic $S$-matrix 
element for $^{4}$He+$^{58}$Ni elastic scattering at 
$E_{\rm in}/A_{\rm P}=26, 85$ and 175~MeV. 
The solid, dashed and dotted lines represent 
the elastic $S$-matrix elements calculated with the DF-TDA model 
at $E_{\rm in}/A_{\rm P}=26, 85$ and 175~MeV, respectively. 
   }
\label{Fig-S-R-He4Ni58E85}
\end{center}
\end{figure}

Figure \ref{Fig-S-R-He4Ni58E85} shows the absolute value of the elastic 
$S$-matrix element as a function of $R$ for 
$^{4}$He+$^{58}$Ni elastic scattering, 
where $R$ is estimated from the angular momentum $L$ between P and T 
with the semi-classical relation $L=R K(\infty)$. 
The solid, dashed and dotted lines correspond to 
the elastic $S$-matrix elements calculated with the DF-TDA model 
at $E_{\rm in}/A_{\rm P}=26, 85$ and 175~MeV, respectively. 
The $^{4}$He scattering becomes more peripheral as $E_{\rm in}/A_{\rm P}$ 
decreases. Particularly at $E_{\rm in}/A_{\rm P}=26$~MeV, the scattering 
is quite peripheral. This is the reason why 
the NAF model well simulates the DF-TDA model for lower energies 
of $E_{\rm in}/A_{\rm P}=20\mbox{--}43$~MeV. 
Eventually, the NAF model is good not only for higher energies of 
$E_{\rm in}/A_{\rm P}=120\mbox{--}175$~MeV but also for lower energies 
of $E_{\rm in}/A_{\rm P}=20\mbox{--}43$~MeV.

\begin{figure}[htbp]
\begin{center}
 \includegraphics[width=0.3\textwidth,clip]{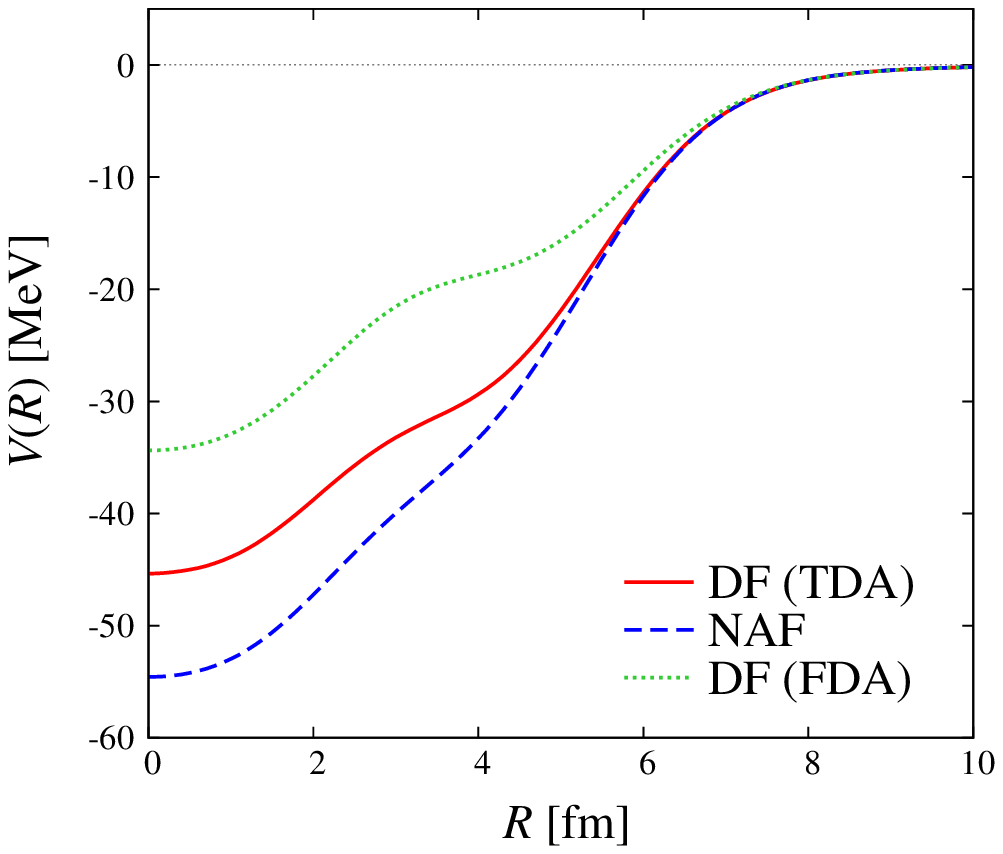}
 \includegraphics[width=0.3\textwidth,clip]{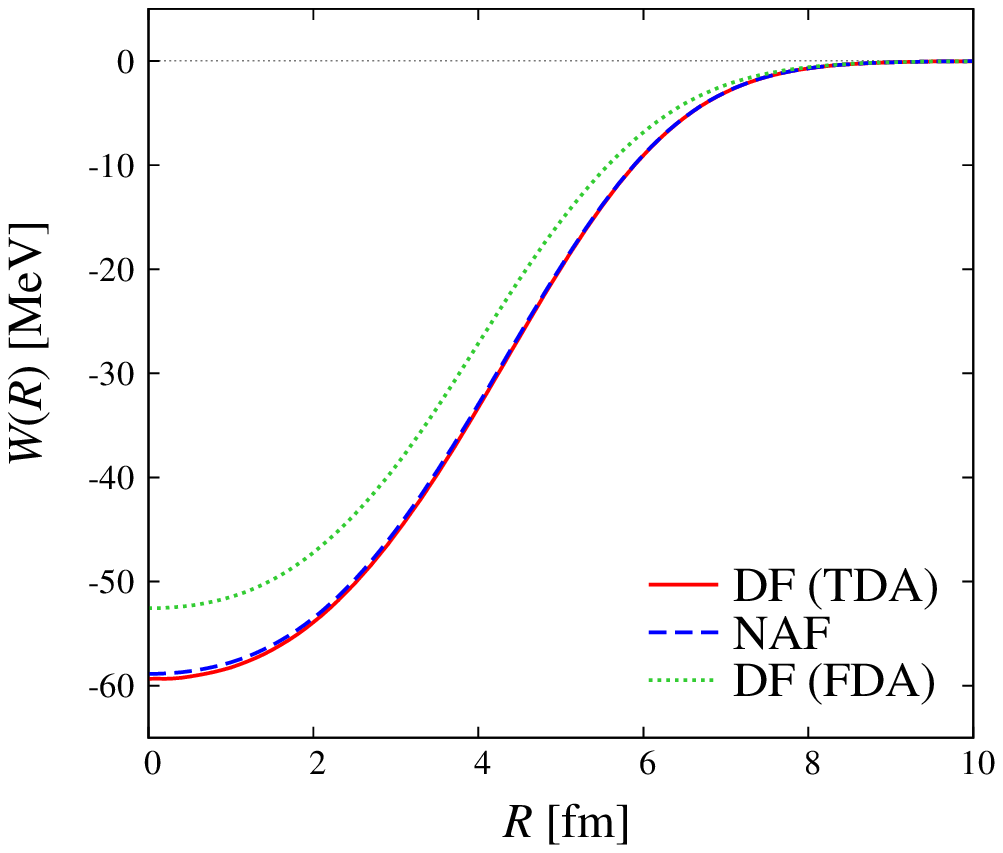}
 \caption{(Color online) 
Optical potentials for $^{4}$He+$^{58}$Ni elastic scattering at 
$E_{\rm in}/A_{\rm P}=175$~MeV. The solid (dotted) line stands 
for the DF-TDA (DF-FDA) potential, 
whereas the dashed line denotes the NAF potential. 
}
 \label{Fig-U-He4Ni58E175}
\end{center}
\end{figure}

Figure \ref{Fig-U-He4Ni58E175} shows the folding potentials 
$U=V+iW$ for $^{4}$He+$^{58}$Ni elastic scattering at 
$E_{\rm in}/A_{\rm P}=175$~MeV. 
The FDA has stronger Pauli-blocking effects than the TDA 
because of $\rho_{\rm P}+\rho_{\rm T} \ge \rho_{\rm P}$. 
As a result of this property, 
the DF-FDA potential (dotted line) is 
less attractive and less absorptive than the DF-TDA potential (solid line). 
The NAF model (dashed line) well simulates the DF-TDA potential 
in $R \ga 5$~fm, as expected. This is the reason why 
at $E_{\rm in}/A_{\rm P}=175$~MeV 
the NAF model well simulates the DF-TDA model for both 
$d\sigma/d\Omega$ and $\sigma_{\rm R}$.

\begin{figure}[htbp]
\begin{center}
 \includegraphics[width=0.3\textwidth,clip]{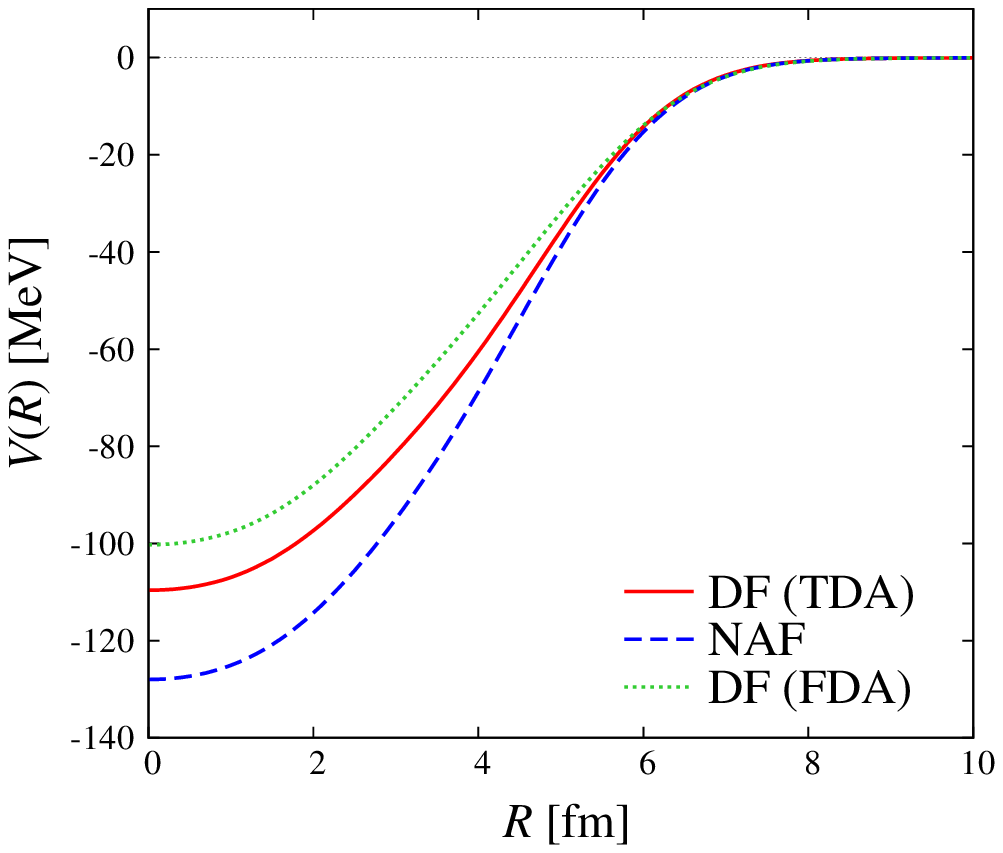}
 \includegraphics[width=0.3\textwidth,clip]{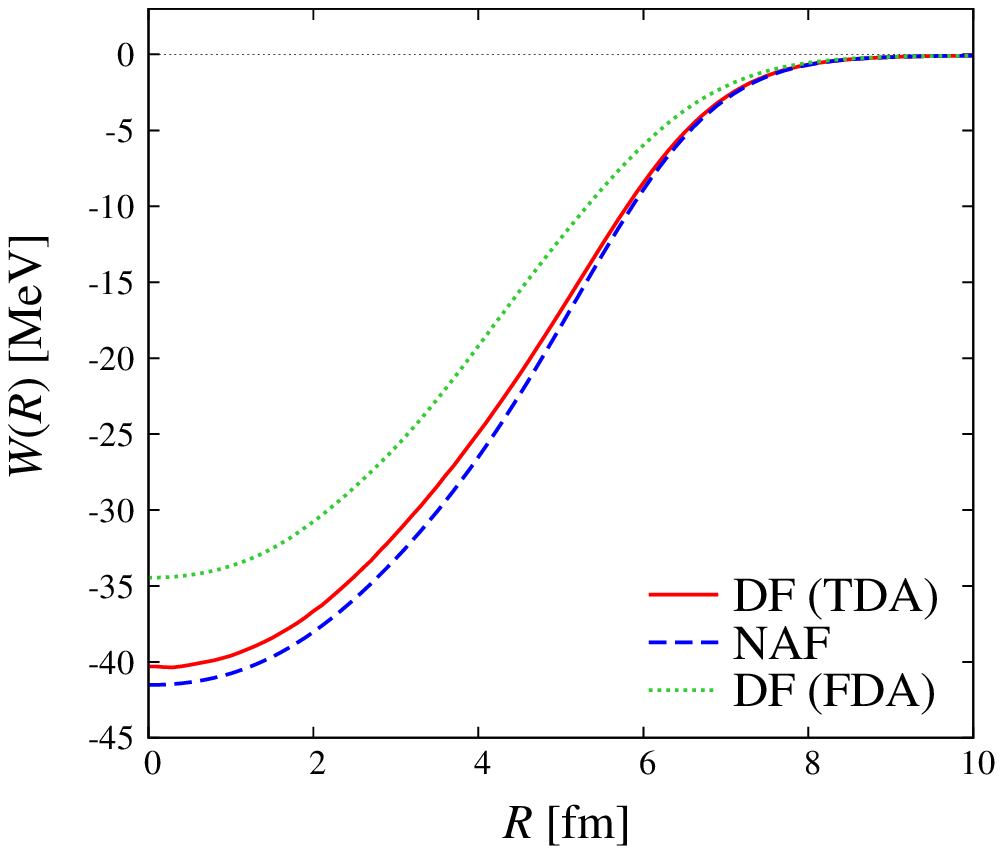}
 \caption{(Color online) 
Optical potentials for $^{4}$He+$^{58}$Ni elastic scattering at 
$E_{\rm in}/A_{\rm P}=85$~MeV. See Fig. \ref{Fig-U-He4Ni58E175} 
for the definition of lines. 
   }
 \label{Fig-U-He4Ni58E85}
\end{center}
\end{figure}

\begin{figure}[htbp]
\begin{center}
 \includegraphics[width=0.3\textwidth,clip]{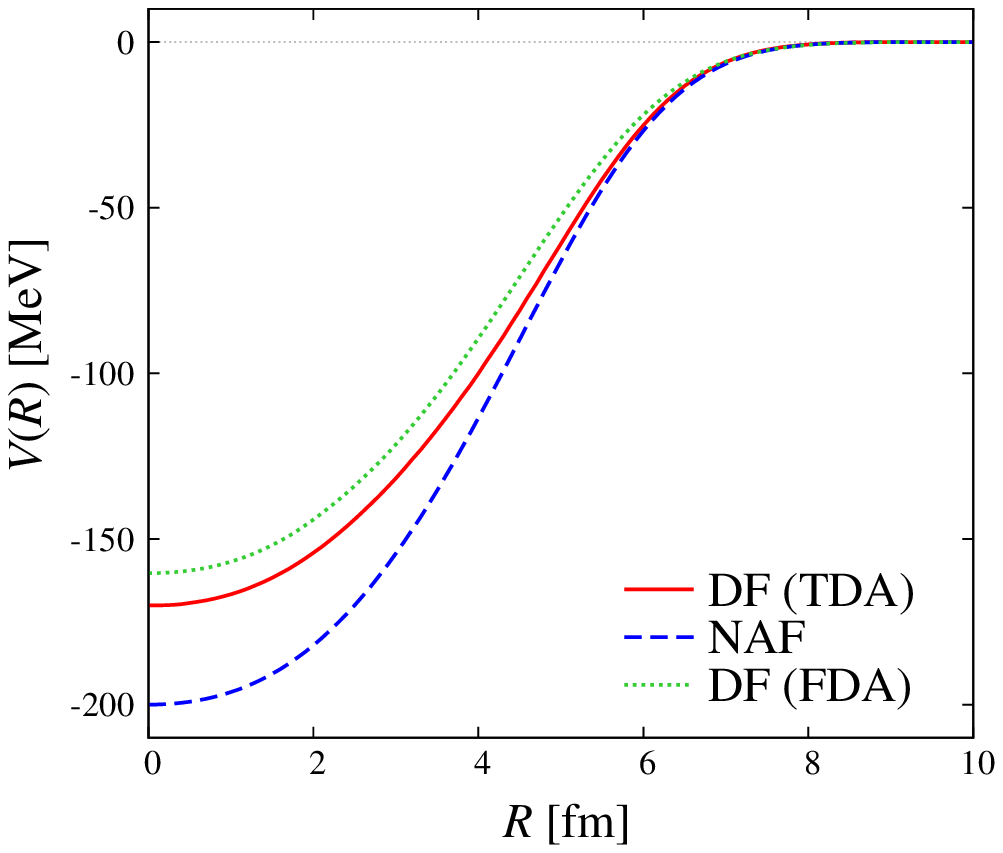}
 \includegraphics[width=0.3\textwidth,clip]{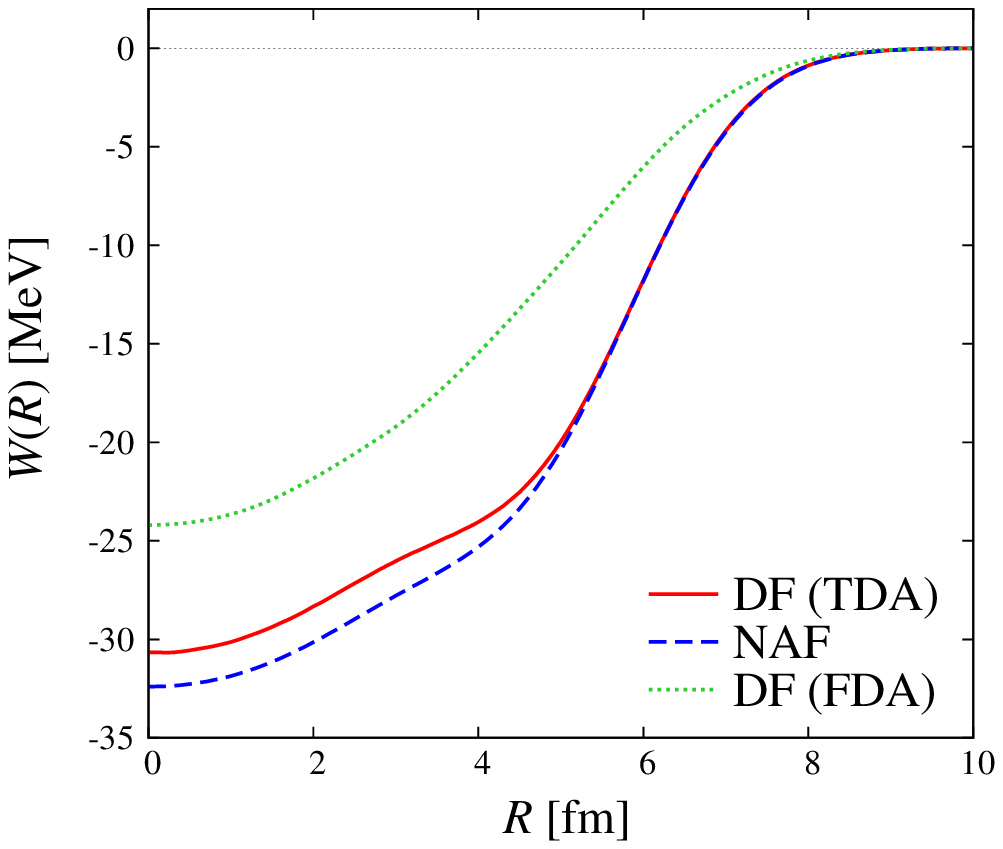}
 \caption{(Color online) 
Optical potentials for $^{4}$He+$^{58}$Ni elastic scattering at 
$E_{\rm in}/A_{\rm P}=26$~MeV. See Fig. \ref{Fig-U-He4Ni58E175} 
for the definition of lines. 
   }
 \label{Fig-U-He4Ni58E26}
\end{center}
\end{figure}

Figures \ref{Fig-U-He4Ni58E85} and \ref{Fig-U-He4Ni58E26} show 
the folding potentials for $^{4}$He+$^{58}$Ni elastic scattering at 
$E_{\rm in}/A_{\rm P}=85$ and 26~MeV, respectively. 
The Pauli-blocking effects due to $\rho_{\rm P}$, which is 
represented by the difference between the DF-TDA and DF-FDA potentials, 
become large as $E_{\rm in}/A_{\rm P}$ decreases, as expected. 
For $E_{\rm in}/A_{\rm P}=26$~MeV, the NAF potential reproduces the DF-TDA 
potential in $R \ga 5$~fm, but the former largely deviates 
from the latter in $R \la 5$~fm. The deviation does not contribute to 
$d\sigma/d\Omega$ and $\sigma_{\rm R}$, since the elastic 
$S$-matrix elements are quite small in $R \la 5$~fm. 
This is the reason why the NAF model is good for lower energies. 
For $E_{\rm in}/A_{\rm P}=85$~MeV, meanwhile, 
the NAF potential is largely deviated from the DF-TDA potential 
in $R \la 5$~fm, whereas the elastic $S$-matrix elements are small 
only in $R \la 3$~fm. The NAF model is thus not good for intermediate 
energies around $E_{\rm in}/A_{\rm P}=85$~MeV.

\begin{figure}[htbp]
\begin{center}
 \includegraphics[width=0.40\textwidth,clip]{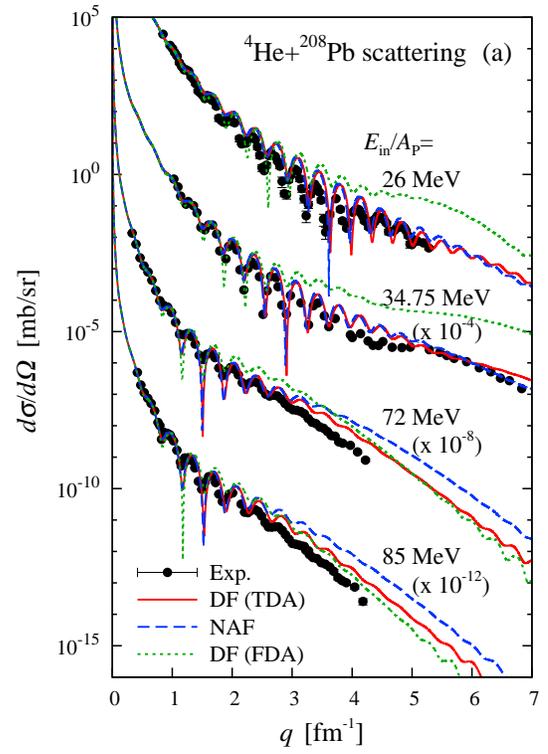}
 \includegraphics[width=0.40\textwidth,clip]{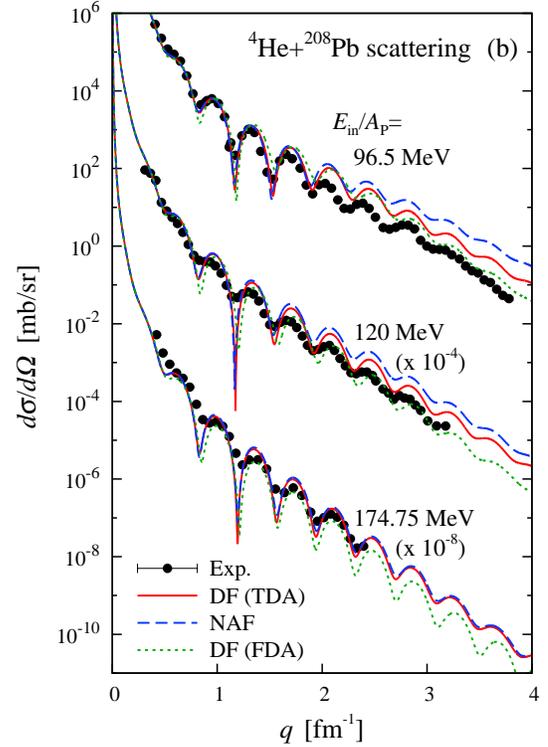}
 \caption{(Color online) 
Differential cross sections as a function of transfer momentum $q$ 
for $^{4}$He+$^{208}$Pb elastic scattering at 
(a) $E_{\rm in}/A_{\rm P}=26\mbox{--}85$~MeV and 
(b) $E_{\rm in}/A_{\rm P}=97\mbox{--}175$~MeV.
The cross section at each $E_{\rm in}/A_{\rm P}$ is multiplied by 
the factor shown in the panel. 
The solid (dotted) line stands 
the results of the DF-TDA (DF-FDA) model, 
whereas the dashed line denotes the results of the NAF model. 
The experimental data are taken from \cite{E26:Hauser:1969,E35:Goldberg:1973,E72-175:Bonin:1985,E97:Uchida:2004}.
}
 \label{Fig-XSEC-He4Pb208}
\end{center}
\end{figure}

Finally we briefly discuss $^{4}$He+$^{208}$Pb elastic scattering. 
Figure \ref{Fig-XSEC-He4Pb208} shows $d\sigma/d\Omega$ as 
a function of $q$ for $^{4}$He+$^{208}$Pb scattering in 
(a) $E_{\rm in}/A_{\rm P}=26\mbox{--}85$~MeV and 
(b) $E_{\rm in}/A_{\rm P}=97\mbox{--}175$~MeV.
The same statement is possible also for $^{208}$Pb target. Namely, 
the DF-TDA model yields better agreement with the experimental data 
\cite {E26:Hauser:1969,E35:Goldberg:1973,E97:Uchida:2004,E72-175:Bonin:1985} 
than the DF-FDA model. The NAF model well simulates the DF-TDA model 
for lower energies around $E_{\rm in}/A_{\rm P}=30$~MeV and also for 
higher energies around $E_{\rm in}/A_{\rm P}=175$~MeV.

\section{Summary}
\label{Summary}

We presented a reliable double-folding (DF) model for $^{4}$He scattering 
from heavier targets such as $^{58}$Ni and $^{208}$Pb 
in a wide range of incident energies from 20 to 200~MeV/nucleon. 
It is the Melbourne $g$-matrix DF model with the target-density approximation (TDA)
, i.e., the DF-TDA model. 
The reliability of the DF-TDA model was shown theoretically with 
the multiple scattering theory and phenomenologically by 
showing that the model reproduces measured $d\sigma/d\Omega$ and 
$\sigma_{\rm R}$. 
The DF-TDA model yields much better agreement with the experimental data than 
the usual DF model with the frozen-density approximation.

We also investigated the reliability of the nucleon-nucleus folding (NAF) 
model in which the nucleon-nucleus (NA) potential is folded with 
the $^{4}$He density. This model is quite practical, since we can use the 
phenomenological NA optical potential instead of the microscopic NA optical 
potential. The NAF model well simulates the DF-TDA model 
for lower energies around $E_{\rm in}/A_{\rm P}=30$~MeV and also 
for higher energies around $E_{\rm in}/A_{\rm P}=175$~MeV.

\vspace*{5mm}

\section*{Acknowledgments}

One of the authors (K. M.) is supported in part by Grant-in-Aid for Scientific Research
(No. 244137) 
from Japan Society for the Promotion of Science (JSPS).

\vspace*{5mm}



\begin{thebibliography}{00}


\bibitem{CDCC-review1}
M.~Kamimura, 
M.~Yahiro, Y.~Iseri, Y.~Sakuragi, H.~Kameyama, and
M.~Kawai, \newblock
Prog.\ Theor.\ Phys.\ Suppl.\ {\bf 89}, 1 (1986).

\bibitem{CDCC-review2}
N.~Austern, 
Y.~Iseri, M.~Kamimura, M.~Kawai, G.~Rawitscher, and
M.~Yahiro, \newblock
Phys.\ Rep.\ {\bf 154}, 125 (1987).

\bibitem{Yahiro:2012tk} 
  M.~Yahiro, K.~Ogata, T.~Matsumoto, and K.~Minomo,
 Prog. Theor. Exp. Phys. {\bf 2012}, 01A206 (2012).


\bibitem{M3Y}
G.~Bertsch, J.~Borysowicz, H.~McManus, and W.~G.~Love,
Nucl. Phys. A {\bf 284}, 399 (1977).

\bibitem{JLM}
J.~-P.~Jeukenne, A.~Lejeune, and C.~Mahaux, Phys. Rev. C{\bf 16}, 80 (1977);\\
J.~-P.~Jeukenne, A.~Lejeune, and C.~Mahaux, Phys. Rep. {\bf 25}, 83 (1976).


\bibitem{Brieva-Rook}
F.~A.~Brieva and J.~R.~Rook, Nucl. Phys. A {\bf 291}, 299 (1977);
{\it ibid.} 291, 317 (1977); {\it ibid.} 297, 206 (1978).

\bibitem{Satchler-1979}
G.~R.~Satchler and W.~G.~Love, Phys. Rep. {\bf 55}, 183 (1979).

\bibitem{Satchler}
G.~R.~Satchler, \lq\lq Direct Nuclear Reactions'',
Oxfrod University Press, (1983).

\bibitem{CEG}
N.~Yamaguchi, S. ~Nagata, and T.~Matsuda, Prog. Theor.
Phys. {\bf 70}, 459 (1983);
N.~Yamaguchi, S.~Nagata, and J.~Michiyama,
Prog. Theor. Phys. {\bf 76}, 1289 (1986).

\bibitem{Rikus-von Geramb}
L.~Rikus, K.~Nakano, and H.~V.~Von Geramb, Nucl. Phys. A {\bf 414}, 413 (1984);
L.~Rikus, and H.~V.~Von Geramb, Nucl. Phys. A {\bf 426}, 496 (1984).

\bibitem{Amos}
K.~Amos, P.~J.~Dortmans, H.~V.~Von Geramb, S.~Karataglidis, and J.~Raynal, in \textit{Advances in Nuclear Physics}, edited by
J.~W.~Negele and E.~Vogt(Plenum, New York, 2000) Vol. 25, p. 275.

\bibitem{CEG07}
T.~Furumoto, Y.~Sakuragi, and Y.~Yamamoto, Phys. Rev. C {\bf 78},
044610 (2008); {\it ibid.},  {\bf 79}, 011601(R) (2009);
{\it ibid.},  {\bf 80}, 044614 (2009).

\bibitem{Saliem}
S.~M.~Saliem and W.~Haider, J. Phys. G {\bf 28}, 1313 (2002).

\bibitem{DFM-standard-form}
B. Sinha, Phys. Rep. {\bf 20}, 1 (1975). \\
B. Sinha and S. A. Moszkowski, Phys. Lett. B{\bf 81}, 289 (1979).


\bibitem{Arellano:1995}
H.~F.~Arellano, F.~A.~Brieva, and W.~G.~Love, 
Phys. Rev. C {\bf 52}, 301 (1995). 

\bibitem{rainbow}
 D.~T.~Khoa, 
 W.~von Oertzen, H.~G.~Bohlen, and S.~Ohkubo,
 J. Phys. G {\bf 34}, R111 (2007).

\bibitem{DFM-standard-form-2}
T. Furumoto, Y. Sakuragi, and Y. Yamamoto, Phys. Rev. C{\bf 82}, 044612 (2010).


\bibitem{Sum12}
T.~Sumi {\it et al.}, 
Phys. Rev. C {\bf 85}, 064613 (2012). 




\bibitem{Minomo:2009ds}
  K.~Minomo, K.~Ogata, M.~Kohno, Y.~R.~Shimizu, and M.~Yahiro,
  J.\ Phys.\ G {\bf 37}, 085011 (2010)
  [arXiv:0911.1184 [nucl-th]].

\bibitem{Hag06}
K. Hagino, T. Takehi, and N. Takigawa,
Phys. Rev. C {\bf 74} (2006), 037601.



\bibitem{Toyokawa:2013uua}
  M.~Toyokawa, K.~Minomo, and M.~Yahiro,
 Phys. Rev. C{\bf 88}, 054602 (2013).





\bibitem{Koning-Delaroche}
A.~J.~Koning and J.~P.~Delaroche, Nucl. Phys. A {\bf 713} 231 (2003).



\bibitem{Dirac1}
S.~Hama, B.~C.~Clark, E.~D.~Cooper, H.~S.~Sherif, and R.~L.~Mercer, Phys. Rev. C {\bf 41}, 2737 (1990).

\bibitem{Dirac2}
E.~D.~Cooper, S.~Hama, B.~C.~Clark, and R.~L.~Mercer, Phys. Rev. C {\bf 47}, 297 (1993).


\bibitem{Perey-Perey}
C.~M.~Perey and F.~G.~Perey, At. Data Nucl. Data Tables {\bf 17}, 1 (1976).




\bibitem{Watson}
K.~M.~Watson, Phys. Rev. {\bf 89}, 575 (1953).

\bibitem{KMT}
A.~K.~Kerman, H.~McManus, and R.~M.~Thaler, Ann.
Phys. {\bf 8}, 551 (1959).


\bibitem{Yahiro-Glauber}
M.~Yahiro, K.~Minomo, K.~Ogata, and M.~Kawai,
Prog. Theor. Phys. {\bf 120}, 767 (2008).



\bibitem{Furumoto:2006ek} 
  T.~Furumoto and Y.~Sakuragi,
  Phys.\ Rev.\ C {\bf 74}, 034606 (2006).



\bibitem{Min11}
K.~Minomo, T.~Sumi, M.~Kimura, K.~Ogata, Y.~R.~Shimizu, and M.~Yahiro,
Phys. Rev. C {\bf 84}, 034602 (2011).

\bibitem{Min12}
K.~Minomo, T.~Sumi, M.~Kimura, K.~Ogata, Y.~R.~Shimizu, and M.~Yahiro,
Phys. Rev. Lett. {\bf 108}, 052503 (2012).

\bibitem{Kimura}
M.~Kimura and H.~Horiuchi, Prog. Theor. Phys. {\bf 111}, 841 (2004).

\bibitem{Kimura1}
M.~Kimura, Phys. Rev. C{\bf 75}, 041302(R) (2007).

\bibitem{GognyD1S}
J.~F.~Berger, M.~Girod, and D.~Gogny,
Comput. Phys. Commun. {\bf 63}, 365 (1991). 





\bibitem{Takeda}
G. Takeda and K. M. Watson, Phys. Rev. {\bf 97}, 1336(1955).

\bibitem{Picklesimer}
A. Picklesimer and R. M. Thaler, Phys. Rev. C{\bf 23}, 42(1981).



\bibitem{C12-density}
H.~de Vries, C.~W.~de Jager, and C.~de Vries,
At. Data Nucl. Data Tables {\bf 36}, 495 (1987).



\bibitem{Singhal}
R.~P.~Singhal, M.~W.~S.~Macauley, and P.~K.~A.~De Witt Huberts, Nucl. Instr. and Meth. {\bf 148}, 113 (1978).


\bibitem{E21:Chang:1976}
H.~H.~Chang, B.~W.~Ridley, T.~H.~Braid, T.~W.~Conlon, E.~F.~Gibson, and N.~S.~P.~King, Nucl. Phys. A {\bf 270}, 413 (1976).


\bibitem{E26:Rebel:1972}
H.~Rebel, R.~L\"{o}hken, G.~W.~Schweimer, G.~Achatz, and G.~Hauser, Z. Phys. {\bf 256}, 258 (1972).


\bibitem{E43:Albinski:1985}
J.~Albi\'nski \textit{et al}., Nucl. Phys. A {\bf 445}, 477 (1985).


\bibitem{E60:Lui:2006}
Y.~-W.~Lui, D.~H.~Youngblood, H.~L.~Clark, Y.~Tokimoto, and B.~John, Phys. Rev. C {\bf 73}, 014314 (2006).



\bibitem{E72-175:Bonin:1985}
B.~Bonin \textit{et al}., Nucl. Phys. A {\bf 445}, 381 (1985).


\bibitem{E97:Nayak:2006}
B.~K.~Nayak \textit{et al}., Phys. Lett. B {\bf 637}, 43 (2006).


\bibitem{E29-48:Ingemarsson:2000}
A.~Ingemarsson \textit{et al}., Nucl. Phys. A {\bf 676}, 3 (2000).


\bibitem{E26:Hauser:1969}
G.~Hauser, R.~L\"{o}hken, H.~Rebel, G.~Schatz, G.~W.Schweimer, and J.~Specht, Nucl. Phys. A {\bf 128}, 81 (1969).


\bibitem{E35:Goldberg:1973}
D.~A.~Goldberg, S.~M.~Smith, H.~G.~Pugh, P.~G.~Roos, and N.~S.~Wall, Phys. Rev. C {\bf 7}, 1938 (1973).


\bibitem{E97:Uchida:2004}
M.~Uchida \textit{et al}., Phys. Rev. C {\bf 69}, 051301(R) (2004).



\end{thebibliography}
\end{document}